\documentclass[]{aa}

\usepackage{mathtools, amssymb, amsmath, glossaries, multirow, booktabs, comment, graphicx, natbib}
\usepackage[normalem]{ulem}
\usepackage[varg]{txfonts}

\begin{document}

   \title{High-$z$ Universe probed via Lensing by QSOs (HULQ) II. \\
Deep GMOS spectroscopy of a QSO lens candidate}

\author{Y. C. Taak \inst{1,2,3} \and M. Im \inst{2,3} \and Y. Kim \inst{4,5} \and M. Hyun \inst{6,2,3} \and I. Paek \inst{2,3}}

   \institute{Physics and Astronomy Department, University of California, Los Angeles, CA 90095-1547, USA
        \and
            SNU Astronomy Research Center, Seoul National University, 1 Gwanak-ro, Gwanak-gu, Seoul 08826, Republic of Korea
        \and
             Astronomy Program, Department of Physics \& Astronomy, Seoul National University, 1 Gwanak-ro, Gwanak-gu, Seoul 08826, Republic of Korea
        \and
            Department of Astronomy and Atmospheric Sciences, College of Natural Sciences, Kyungpook National University, Daegu 41566, Republic of Korea
        \and
            Kavli Institute for Astronomy and Astrophysics, Peking University, Beijing 100871, People's Republic of China
        \and
            Korea Astronomy and Space Science Institute, Daejeon 34055, Republic of Korea\\
        \email{yctaak@astro.ucla.edu,myungshin.im@gmail.com}
             }

   \date{12 July 2022}

\abstract{Galaxies and their central supermassive black holes are known to coevolve, but the physical background for this is unknown as of yet. The High-$z$ Universe probed via Lensing by QSOs (HULQ) project aims to investigate this coevolution by using quasi-stellar object (QSO) host galaxies acting as gravitational lenses (QSO lenses). We present the results of the spectroscopic observation of the first QSO lens candidate from the HULQ project, HULQ J0002+0239, which consists of a QSO host galaxy at $z_{\rm d} = 1.455$ and four seemingly lensed objects in a cross-like configuration. Deep optical spectra of two of the possibly lensed objects with $z \sim 24.5$ mag were obtained with the Gemini Multi-Object Spectrograph on the Gemini North Telescope. Their spectra reveal that the objects are newly discovered galaxies at $z=0.29$ and $z=1.11$, and we conclude that HULQ J0002+0239 is not a QSO lens. Our QSO lens search results are so far in agreement with the predicted number of QSO lenses, and we discuss how the future investigation of additional QSO lens candidates could tell us more about the evolution of the black hole mass and host galaxy scaling relations. }


\titlerunning{HULQ II. Deep GMOS spectroscopy of a QSO lens candidate}
\authorrunning{Y. C. Taak et al.}
   \maketitle
%
%-------------------------------------------------------------------

\section{Introduction} \label{sec:intro}
It is now accepted that supermassive black holes (SMBHs) and their host galaxies coevolve to some extent. Correlations between properties of SMBHs and their hosts are presented as evidence, most commonly the relation between the mass of a SMBH and the velocity dispersion of its host galaxy \citep[$M_{\rm BH} - \sigma_*$ relation;][]{Ferrarese+00,Gebhardt+00a,Kormendy+13}. Yet the evolution of such correlations over cosmic time, which will allow us to address the coevolution mechanism more directly, is not fully understood \citep{ShenY+15,DingX+17}. 

To study these correlations in the early universe, observations of SMBHs at high redshifts are required. Quasi-stellar objects (QSOs) are valuable probes in this manner, since their luminous nature allows us to discover them at all epochs after reionization \citep{Schneider+02,FanX+06,Schneider+10,KimY+15,KimY+19,KimY+20,Banados+16,JeonY+16,JeonY+17,JiangL+16, Paris+17,Paris+18,Lyke+20,ShinS+20,WangF+21}. In addition, methods for estimating the masses of their central engines have been studied in great detail \citep{Peterson+04,Vestergaard+06,Bentz+09,Vestergaard+09,KimD+10,KimD+15,JunH+15}. Reverberation mapping is the most accurate method, but it requires spectroscopic monitoring programs, which are extremely time-consuming. The single-epoch method is the preferred alternative for large samples of QSOs due to its simple observation requirement. The issue is studying their host galaxies, since it becomes difficult to even observe the hosts starting at $z \sim 1$, not simply because their angular sizes decrease to subarcsecond levels \citep{ImM+95,PengC+06,Decarli+10,Merloni+10}, but also due to the sheer brightness of the QSOs themselves \citep{Gebhardt+00b}. 

Strong gravitational lensing occurs when a background source is located sufficiently adjacent, angle-wise, to a massive lens (hereafter deflectors, to avoid confusion with lens systems) at a lower redshift. If QSO host galaxies are found to act as strong lenses (hereafter QSO lenses; \citealt{Courbin+10,Courbin+12,Meyer+19}), they function as unique probes to their mass distributions using lens modeling. This enables us to investigate the redshift evolution of the correlations between properties of a SMBH and their hosts, and in turn the SMBH-galaxy coevolution scenario. For example, the black hole mass ($M_{\rm BH}$) -- host stellar velocity dispersion ($\sigma_*$) relation can be explored at various redshifts, where $M_{\rm BH}$ is obtained from the QSO spectral fitting and $\sigma$ from the lens analysis with a plausible assumption for the mass profile (e.g., isothermal spherical mass distribution). 

The High-$z$ Universe probed via Lensing by QSOs (HULQ) project (\citealt{TaakY+20}; hereafter Paper I) aims to utilize QSO lenses to study the coevolution mechanism. It is designed to search for QSO lenses from archival imaging data, confirm the candidates as lens systems using spectroscopy, and analyze the systems by modeling the lensed images with mass models for the deflectors and deduce their mass distributions. Paper I presents the expected number of QSO lenses for various concurrent and future imaging surveys, demonstrating that numerous QSO lenses are to be found, and implying that a direct investigation of the cosmic evolution of SMBH-galaxy correlations with QSO lenses is feasible. Also, it is shown that even if the redshift distribution of the QSO lenses is narrow, it is possible to deduce the correlation evolution by employing distributions of the properties of QSO lenses, such as their Einstein radii ($\theta_{\rm Ein}$). 

In this paper, we present our QSO search in the Hyper Suprime-Cam Survey Strategic Program (HSC-SSP) Wide Survey (HSC/Wide; \citealt{Aihara+18}) imaging data, and spectroscopic results of a promising QSO lens candidate, HULQ J000219.61+023916.0 (hereafter HULQ J0002+0239). We obtained spectra of faint objects surrounding the central QSO to confirm its QSO lens nature, and we conclude that it is not a QSO lens. Using the analysis results, we discuss its implications for the HULQ project. 

All magnitudes are in the AB system, unless denoted otherwise \citep{Oke+83}. The standard cosmological model with cold dark matter and a cosmological constant with $H_0 = 70$ km s$^{-1}$ Mpc$^{-1}$, $\Omega_{\rm M} = 0.3$, and $\Omega_\Lambda = 0.7$ is used, which is supported by observational studies in the past decades \citep{ImM+97,Planck+20}.

\section{QSO lens search} \label{sec:search}
According to the calculations of Paper I, among completed and ongoing surveys, HSC/Wide is expected to harbor the largest number of QSO lenses among concurrent surveys ($\sim$440) due to its superior depth ($\sim$26 mag in $i$) and area ($\sim$1400 deg$^2$). In addition, its excellent mean seeing ($0\farcs6$ in $i$) suggests that a significant fraction of these QSO lenses ($\sim$80) have lensed images that are not hidden within the QSO shot noise, and thus they are detectable. Based on these results, a survey for QSO lenses in imaging surveys is being conducted. We describe this search procedure in the following paragraphs. 

The QSO sample used for the QSO lens search is obtained from the Sloan Digital Sky Survey (SDSS) Data Release 14 (DR14) Quasar Catalog \citep{Paris+18}. The ``CORE'' sample has a magnitude cut of $g<22$ or $r<22$ mag, and these QSOs are located at $0.9 < z < 2.2$. Among the $\sim$530,000 QSOs, we select those that are covered by the $g$, $r$, $i$, and $z$ filters of HSC/Wide in their second public data release \citep[PDR2,][]{Aihara+19}, the most recent data release at the time of selection. For PDR2, the depth of the $i$-filter imaging data reaches 26.2 mag, and the number of QSOs observed with the $i$ filter is $\sim$35,000. 

We gather imaging data for the $\sim$35,000 QSOs observed by the four filters from the HSC Data Archive System (DAS) centered on the QSO, and subtract the QSO and its host galaxy for each image. For accurate fitting, we first mask out other objects by running \texttt{SExtractor} \citep{Bertin+96} on the images with minimum deblending (small \texttt{DEBLEND\_NTHRESH} (16) and large \texttt{DEBLEND\_MINCONT} (0.01) values). Then we use \texttt{Galfit} \citep{PengC+02,PengC+10} to fit the point spread function (PSF) + S\'ersic model and subtract it. We only constrain the positions of the two objects to the center of the image. The PSF for each QSO is obtained from the HSC PSF Picker\footnote{\url{https://hsc-release.mtk.nao.ac.jp/psf/pdr2/}} by inputting its coordinates. In the case that a PSF is not available at the QSO position, we use a median-combined PSF created using all other PSFs for the filter. In some cases the host galaxy is barely detectable, and the S\'ersic profile is unnecessary; to account for this, we also use a PSF-only model and adopt the model-subtracted residual image that provides a better fit between the two models by comparing the reduced $\chi^2$ for the residual images. 

We then run \texttt{SExtractor} again on the residual image to detect objects that may be lensed features. \texttt{SExtractor} is first executed for the $i$-filter image, and then it is run in dual mode for the other three images if they exist, with the $i$-filter image as the detection image. We look for objects that are located close to the QSOs angle-wise (1$\farcs5 < \theta <  5\arcsec$); the lower limit on the angular distance from the QSO is imposed to evade spurious detections caused by the QSO host galaxy, and also to avoid overlap with the QSO lens candidates discovered by \citet{Courbin+10}, \citet{Courbin+12}, and \citet{Meyer+19} using SDSS spectra. Our choice of $1\farcs5$ for the lower limit for $\theta$ is unfortunate in that a large fraction of QSO lenses will be missed based on the Einstein radius distribution of QSO lenses (Figure 13 in Paper I, reproduced here as Figure 8); only 7.0\% and 5.8\% remain after the $\theta$ cut. However, in this study we put more emphasis on discovering QSO lenses that are relatively easy to detect and confirm, rather than providing a complete sample of QSO lens candidates. This also effectively works around the effects of dust extinction by the host galaxy, which may alter the colors of the lensed images significantly. 

Next, we select objects that satisfy the following criteria: $i<25$ mag, to select more likely QSO lens candidates; identical colors ($g-i$, $r-i$, and $i-z$) that agree within 2$\sigma$ errors, to ensure they are from the same source; and a sufficiently large position angle range measured from the QSO, to check for potential lensed image configurations. For instance, if all the objects satisfying the above criteria are located in a single quadrant assuming that the QSO is at the center, then this configuration cannot be explained by a deflector located at the QSO position, and thus it is highly unlikely to be a QSO lens. There are caveats with this search methodology; for sources lying close to the deflector caustic, it is possible that the brighter images are above the image depth and located adjacent to each other, while the remaining images are below the detection limit. These systems cannot be identified using imaging data alone with our algorithm. However, as mentioned above, we focus on the discovery of QSO lenses instead of exploring all possibilities. We use $g-i$ instead of $g-r$ since we are considering all QSOs with $i$-filter imaging data, but not necessarily all four at the same time; even when imaging data of the QSO with a certain filter is missing, it can be selected as a candidate as long as the colors involving the other filters satisfy the criteria. The number of candidates after this stage is 79.

Finally, we visually inspect and grade the remaining QSO lens candidates. Candidates with image configurations and shapes that cannot be explained by gravitational lensing (i.e., objects that are extended radially) or those that are obvious non-candidates (i.e., residual galaxy features) are rejected. We classify the remaining candidates into two categories; Grade A candidates, which have objects that are tangentially extended (more likely to be lensed) and configured such that they can be reproduced by a lens model, and Grade B candidates, which have two point sources that are usually reproducible with a simple lens model, but do not have any further justification as a lens system. Because it is possible that two objects with identical colors are located close to a QSO by coincidence, Grade B candidates are likely to be spurious candidates. The final number of candidates after visual inspection is six and 34 for Grade A and B candidates, respectively. We also classify the candidates using the inferred $\theta_{\rm Ein}$ into large- and small-$\theta_{\rm Ein}$ candidates by applying a cut at $3\farcs5$. The number of candidates is 2+26 Grade A and B candidates for the large-$\theta_{\rm Ein}$ bin, and 4+8 for the small-$\theta_{\rm Ein}$ bin.

\section{HULQ J0002+0239, a QSO lens candidate}
Among the candidates, HULQ J0002+0239 stood out as the most promising large-$\theta_{\rm Ein}$ Grade A candidate. We describe the properties of the QSO lens in the following subsections.

\subsection{SDSS J0002+0239}
\begin{figure}
\centering
\includegraphics[width=0.4\textwidth]{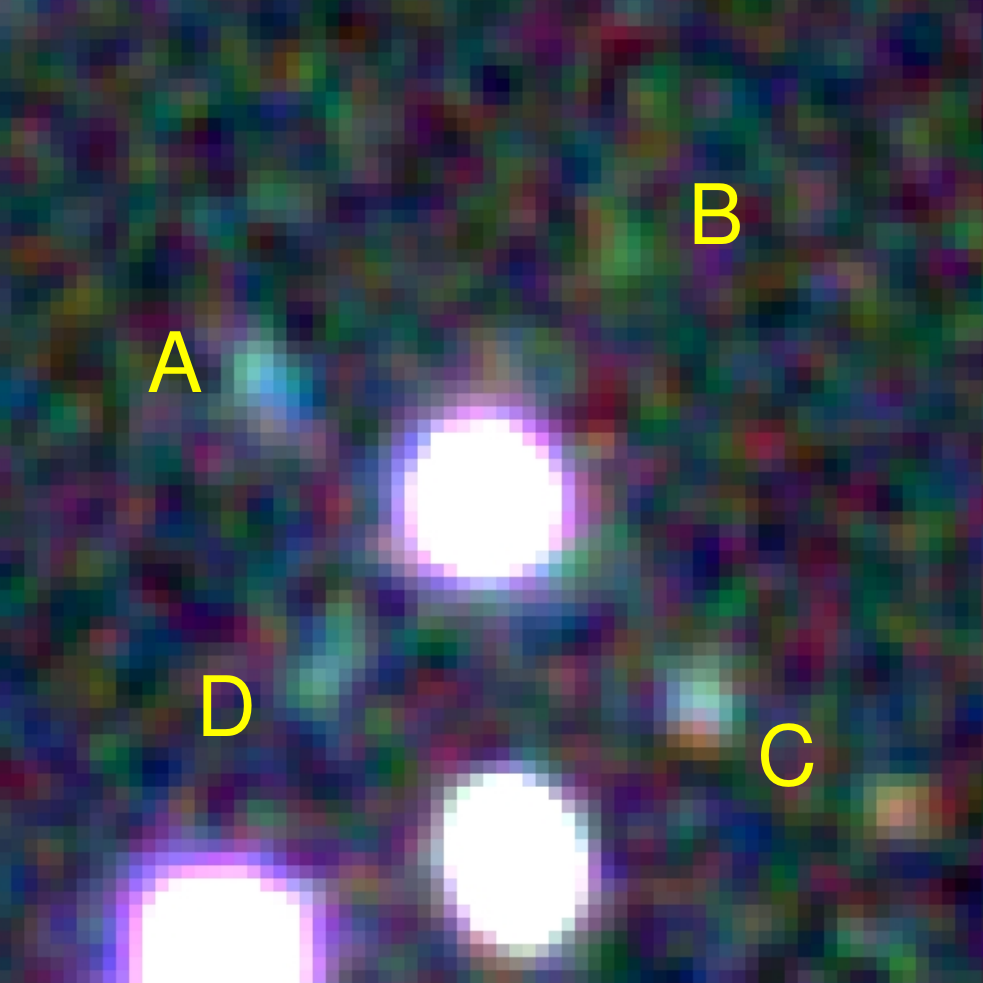}
\includegraphics[width=0.4\textwidth]{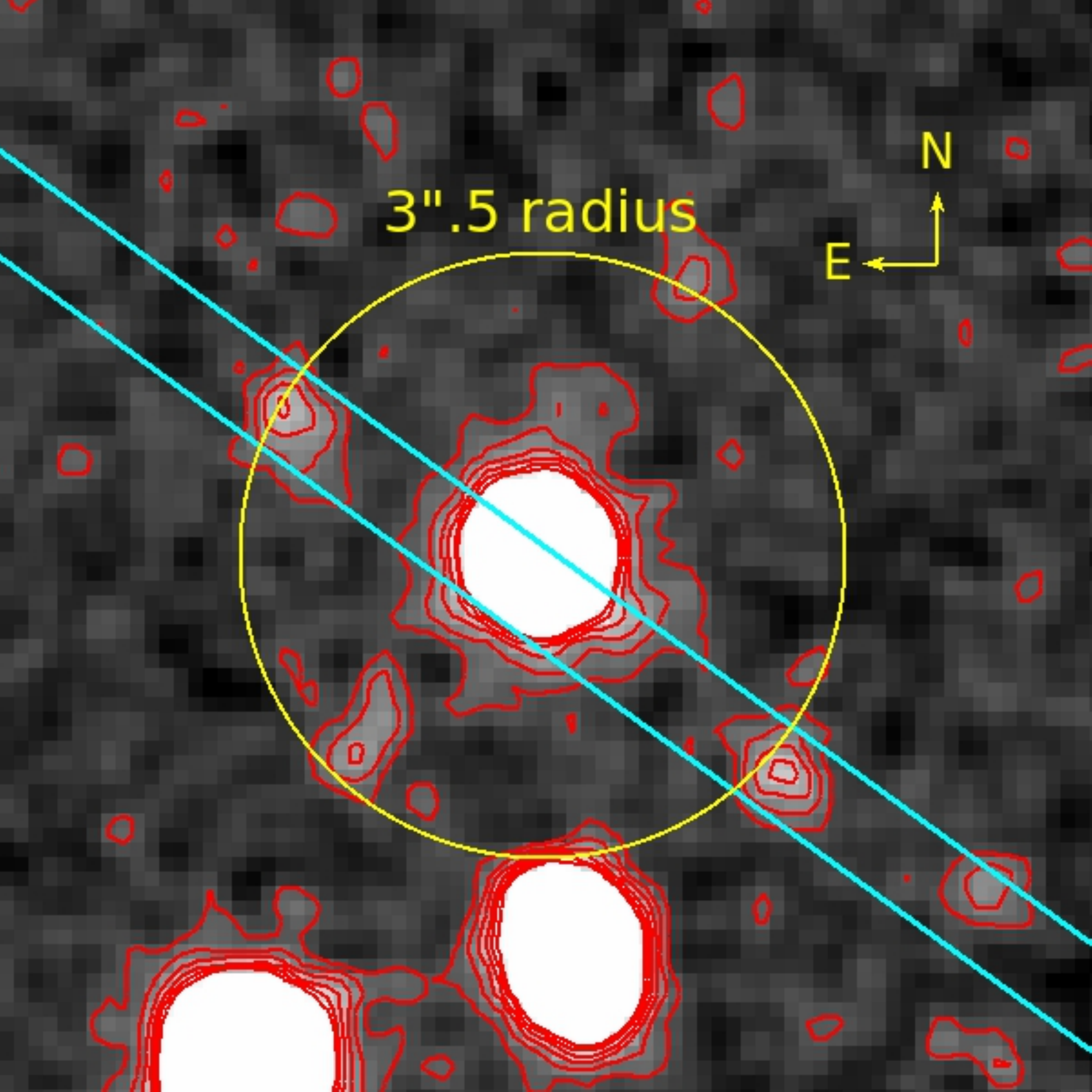}
\caption{Imaging data of HULQ J0002+0239. (Left) HSC/Wide PDR2 $riz$ color image of HULQ J0002+0239. The four objects that are lensed-image candidates are denoted from A to D. (Right) HSC/Wide PDR2 $i$-filter image of HULQ J0002+0239. Red contours indicate isophotes, and they are shown for clarity. The yellow circle is of radius 3$\farcs$5 centered on the QSO, and the cyan box demonstrates the position of the 1$\arcsec$-wide slit used for the GMOS spectroscopy.
}
\label{fig:HSC}
\end{figure}

\begin{figure}
\makebox[0.5\textwidth][c]{\includegraphics[width=\columnwidth]{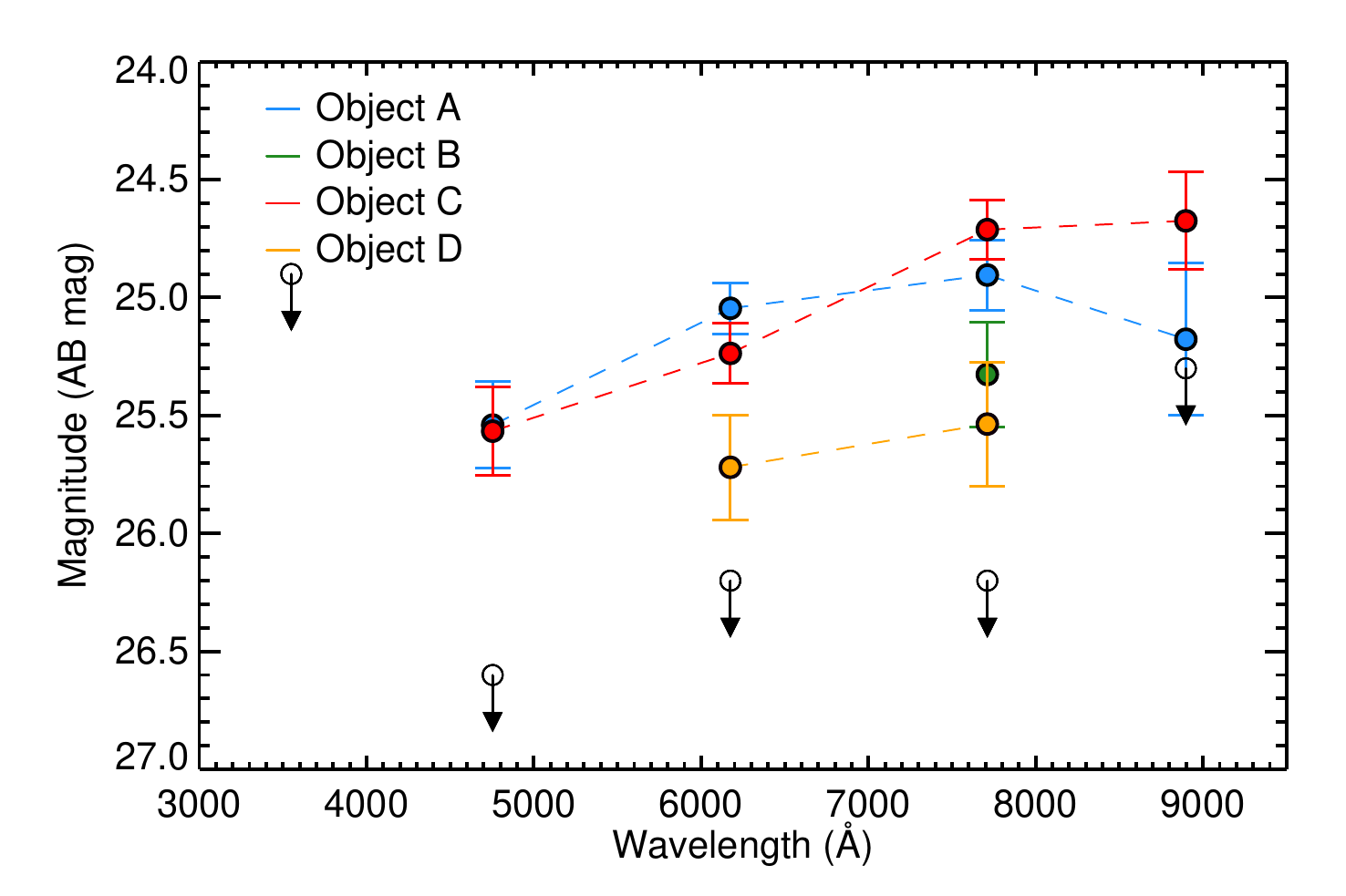}}
\caption{SEDs of the four objects, A, B, C, and D, from the Magellan IMACS ($u'$) and HSC/Wide ($griz$) photometry. Blue, green, red, and yellow symbols are for objects A, B, C, and D, respectively. Circles with black outlines indicate detections with the 1$\sigma$ magnitude errors, while open circles with arrows indicate the image or survey depths. 
}
\label{fig:SED}
\end{figure}

The central QSO, SDSS J0002+0239, is a type 1 QSO at $z=1.455$, and its coordinates are $\alpha =00^{\rm h}02^{\rm m}19.61^{\rm s}$ and $\delta =+02^{\circ}39\arcmin16.0\arcsec$. Figure \ref{fig:HSC} is a cutout of the HSC $riz$ imaging data, which shows four faint objects ($r = 24.6$ -- 26.2 mag) located along a circle of $\sim$3$\farcs5$ radius centered on the QSO, and in a cross-like configuration, which implies a similar $\theta_{\rm Ein}$. We name these four objects from A to D, starting with the northeast object and progressing clockwise. Photometric measurements were conducted with \texttt{SExtractor} \citep{Bertin+96}; the results are listed in Table \ref{tbl:phot}, and the spectral energy distributions (SEDs) of the objects are shown in Figure \ref{fig:SED}. Although the $r$-band photometry somewhat negates the hypothesis that the four objects have similar SEDs, the photometry from other bands supports the possibility of them originating from a single source. In addition, the clear detections in the $g$ filter indicate that these objects are not $g$ dropouts, suggesting that their redshifts are $z_{\rm s} \lesssim 4$ (e.g., see \citealt{KangE+09}).

\citet{Rakshit+20} measured the mass of the SMBH of the QSO from its SDSS spectrum by employing the single-epoch method, which combines the width of the Mg II line and the continuum luminosity at 3000\AA{} via the virial theorem. Their estimate is $M_{\rm BH}=10^{9.14} M_{\odot}$, and in translating this to the (stellar) bulge mass using the local scaling relation for elliptical galaxies given as Equation (10) of \citet{Kormendy+13}, the mass of the bulge is expected to be $M_{\rm bulge} = 10^{11.34} M_{\odot}$.

\subsection{IMACS follow-up imaging}
To solidify the target as a QSO lens, we observed this target with the Inamori Magellan Areal Camera and Spectrograph (IMACS; \citealt{Dressler+11}) on the Magellan Baade 6.5-m Telescope at Las Campanas Observatory, Chile, on 2019 August 31. The f/2 camera with a new Sloan $u'$ filter with improved transmission was used, and the object was observed with single exposure times of 600s in nine frames. One frame was discarded due to a tracking error, resulting in a 5$\sigma$ limiting magnitude for point sources of 24.9 mag for the combined image of 4800s, which is shown in Figure \ref{fig:SED}. This is much shallower than the expected depth of 27.0 mag, and was due to technical issues and poor weather conditions. All four objects have either no or marginal detections, which supports the hypothesis that the objects originate from the same source to some extent. 

\subsection{Gravitational lens modeling}
\begin{figure}
\centering
\includegraphics[width=\columnwidth]{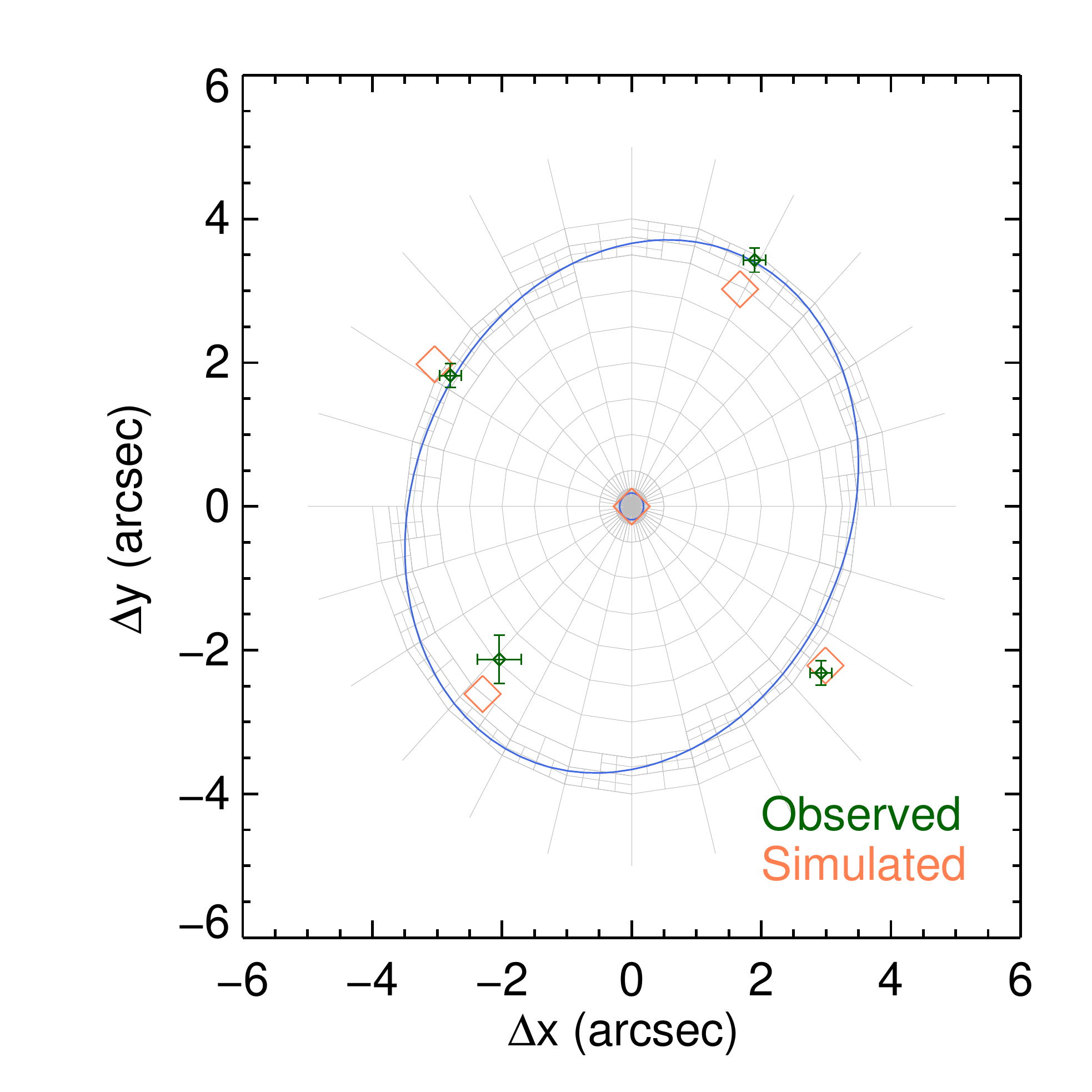}
\caption{\texttt{lensmodel} results for the simulated QSO lens on the image plane. Gray lines are for the grid used for the simulation, and blue lines show the critical curves for the lens system. Green and pink diamonds indicate the observed and simulated positions of the objects, respectively. 
}
\label{fig:lensmodel}
\end{figure}

Assuming that HULQ J0002+0239 is a QSO lens, we checked whether the positions and flux ratios of the ``lensed images'' could be reproduced. Using \texttt{lensmodel} \citep{Keeton11}, a gravitational lens analysis software, a simple model with a singular isothermal ellipsoid as the deflector and a point source recreates the image configuration as shown in Figure \ref{fig:lensmodel}. Five parameters are used for the model: the Einstein radius, ellipticity, and position angle of the deflector, and the relative position of the source with respect to the deflector centroid. With 11 constraints (eight for the four object positions and three for the flux ratios) and five model parameters, the fit gives $\chi^2_{\rm red} = 25.1/6 = 4.18$, further justifying it as a promising QSO lens candidate. A positional error of one pixel was assigned for each of the images with the exception of object D, whose position is quite uncertain, and it was assigned an error of twice the pixel size. Details of the model and the resulting simulated flux ratios are listed in Tables \ref{tbl:lensmodel} and \ref{tbl:lensmodel2}, respectively. 

\begin{figure}
\centering
\includegraphics[width=\columnwidth]{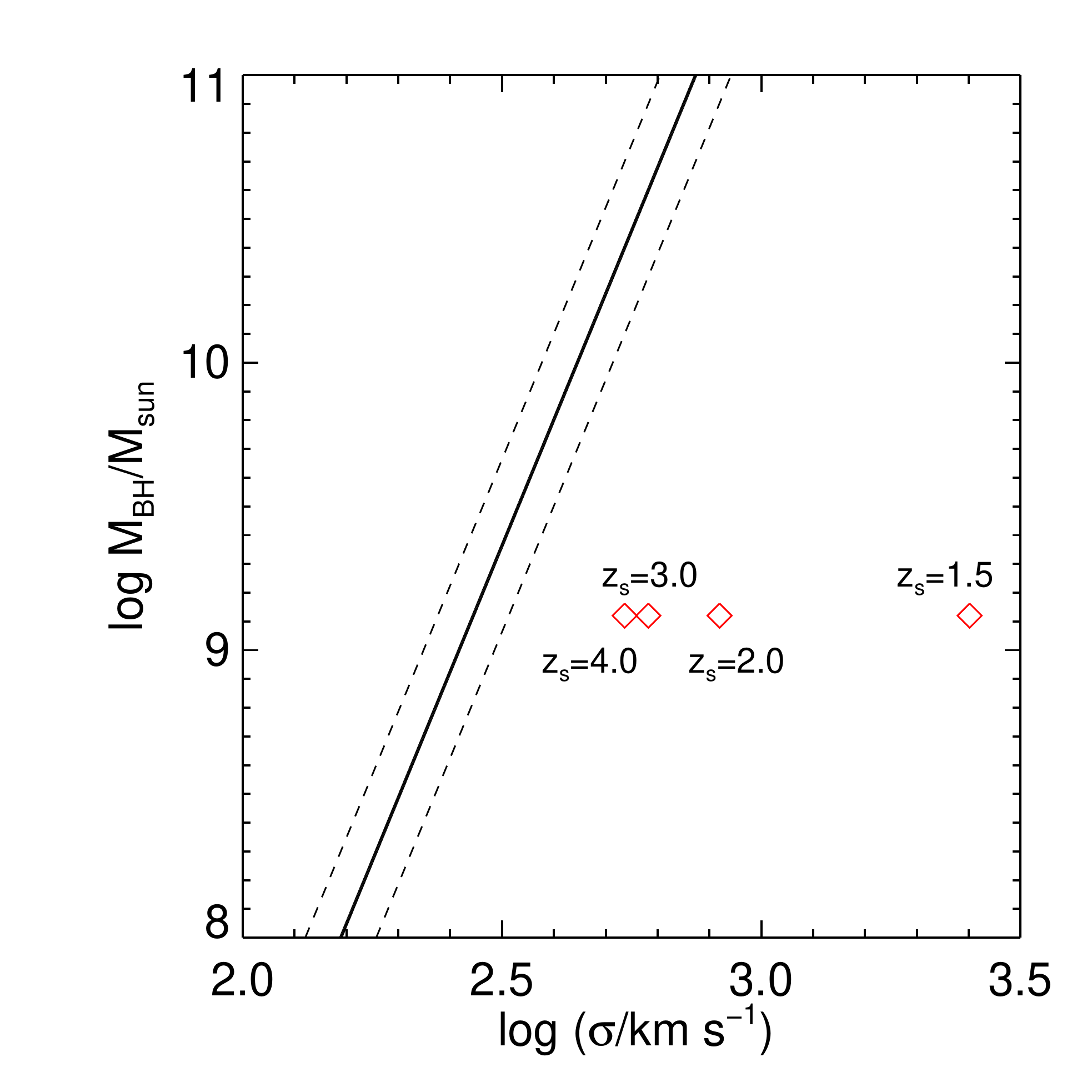}
\caption{$M_{\rm BH} - \sigma_*$ diagram. The local relation and its errors from \citet{Kormendy+13} are shown as black solid and dashed lines, respectively. Red diamonds indicate the positions of SDSS J0002+0239 for four different values of $z_{\rm s}$. 
}
\label{fig:mhalo}
\end{figure}

We used $\theta_{\rm Ein}$ to calculate various properties of the deflector host galaxy; $\sigma$, the velocity dispersion of the host galaxy, was calculated as 
\begin{equation}
\begin{aligned}
\dfrac{\sigma}{c} = \Big( \dfrac{\theta_{\rm Ein}}{4\pi} \: \dfrac{D_{\rm os}}{D_{\rm ds}} \Big)^{1/2}, \label{eq:sigma}\\
\end{aligned}
\end{equation}
where $c$ is the speed of light, and $D_{\rm os}$ and $D_{\rm ds}$ are the angular diameter distances from the observer to the source and from the deflector to the source, respectively. It is important to note that the velocity dispersion from lens modeling and the stellar velocity dispersion agree very well; the ratio of the two ($f\equiv\sigma/\sigma_*$) for the Sloan Lens ACS sample is $1.010\pm0.017$ with 0.065 scatter \citep{Treu+06}. Figure \ref{fig:mhalo} plots $M_{\rm BH}$ and the velocity dispersion of the QSO host for several values of $z_{\rm s}$, along with the local $M_{\rm BH} - \sigma_*$ relation for ellipticals from \citet{Kormendy+13}. We can see that Equation \ref{eq:sigma} and the $z_{\rm s} \lesssim 4$ constraint combined gives $\sigma \gtrsim 550$ km s$^{-1}$. This is substantially larger than the seeming limit of $\sigma \sim 400$ km s$^{-1}$ for local elliptical galaxies \citep{Saulder+15,Yildirim+17}; however, at higher redshifts of $1.5 < z < 2.5$, some compact, quiescent galaxies with $\sigma$ up to 450--500 km s$^{-1}$ have been found \citep{vanDokkum+09,Belli+17}. Thus, if HULQ J0002+0239 is indeed a QSO lens, the QSO host galaxy could be a very compact and massive quiescent galaxy, or it may be a member of a larger galaxy group. 

Regardless of $z_{\rm s}$, $\sigma$ of the host galaxy of SDSS J0002+0239 inferred from the QSO lens assumption lies significantly below the local relation; this is in contrast to the results from the most recent studies (e.g., \citealt{DingX+21}), but is in line with \citet{IshinoT+20}. This disagreement implies that the system may not be a QSO lens. Also, \citet{Reines+15} suggest that active galactic nucleus (AGN) host galaxies have intrinsically lower stellar masses than their quiescent-galaxy counterparts by more than an order of magnitude, which may also be the reason for this discrepancy.

\section{GMOS spectroscopy} \label{sec:obs}
HULQ J0002+0239 was observed in longslit mode with the Gemini Multi-Object Spectrograph (GMOS; \citealt{Hook+04}) on the 8-m Gemini North Telescope at Mauna Kea, Hawaii, on 2020 September 16 and 17 (Program ID: GN-2020B-Q-213). The purpose of this observation is to obtain spectra of the objects and confirm whether they are lensed images from the same source. As can be seen in Figure \ref{fig:HSC}, the 1$\arcsec$-wide slit was positioned along A and C, the two brighter objects, with the QSO located slightly off-center. Due to the faintness of the objects, the largest period grating, R150\_G5308, and four-by-four binning were used to maximize the signal-to-noise ratio (S/N), resulting in a spectral resolution of $R \sim 315.$ Since the nature and redshift of the ``source'' are unknown, we did not use a blocking filter to cover the widest wavelength range possible. To avoid the CCD gaps, observations were conducted with three central wavelengths of 675, 700 and 725nm. Six spectra were obtained for each central wavelength with single exposure times of 990s, for a total of 17820s ($\sim$5 hr). 

The data were reduced following standard procedures using the Gemini AstroConda distribution for PyRAF. First, the bias was subtracted and flat-fielding was applied. The wavelength solution created from the CuAr arc lines was applied. After sky subtraction, the flux calibration was administered using a standard star (BD+28 4211). Finally, the 1D spectra of objects A and C were extracted from each image using five-pixel apertures, equivalent to $1\farcs6$, and then median-combined to generate the final 1D spectra.

\section{Spectroscopic results}

\subsection{Spectroscopic redshifts of objects A and C}
\begin{figure}
\makebox[\textwidth][c]{\includegraphics[width=1.2\textwidth]{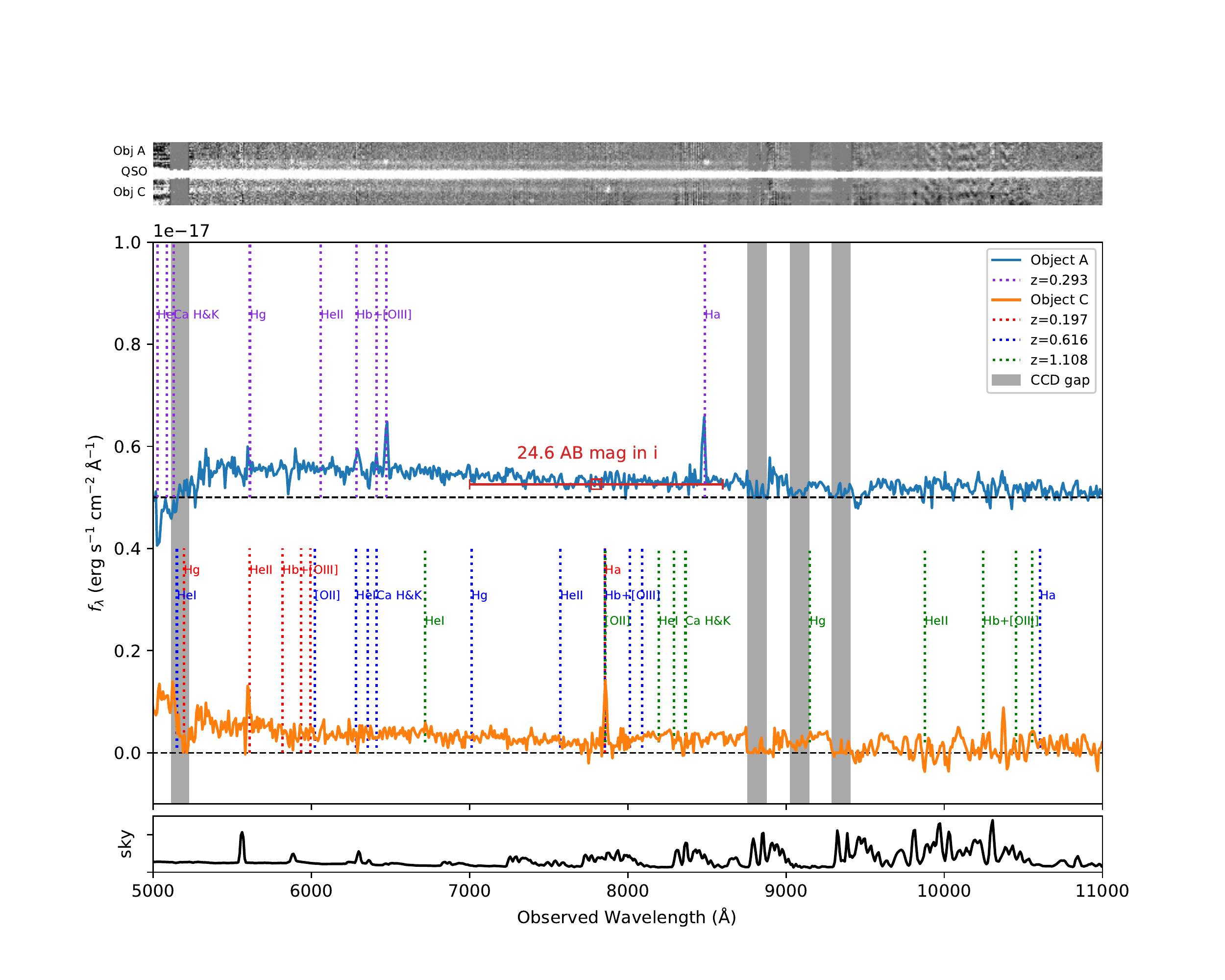}}
\caption{Gemini-N/GMOS spectra of HULQ J0002+0239.
Top panel: 2D spectra of the central QSO and the two objects. 
Center panel: Spectra of the two objects, A and C. The spectra of objects A and C are shown in blue and orange, respectively. The spectra are offset by $0.5\times10^{-17}$ erg s$^{-1}$ cm$^{-2}$ \AA{}$^{-1}$ for clarity, and the horizontal black dashed lines represent zero flux for each spectrum. The HSC $i$-band magnitude of object A is shown with a red square, with the horizontal error bars indicating the bandwidth. For object A, wavelengths of emission lines from an object at $z=0.293$ are indicated with vertical purple dotted lines. For object C, wavelengths of emission lines from objects at $z=$ 0.197, 0.616, and 1.108 are shown with vertical red, blue, and green dotted lines, respectively. These redshifts correspond to the redshifts of the source when the single strong emission line at $\sim$7860\AA{} corresponds to H$\alpha$, H$\beta$, and [O II], respectively. Vertical gray regions indicate the positions of the CCD gaps.
Bottom panel: Sky emission in arbitrary units. 
}
\label{fig:spec}
\end{figure}

Figure \ref{fig:spec} shows the 2D spectrum of the target and the 1D spectra of A and C. For object A, several emission lines are visible; their wavelengths suggest that they originate from a source at $z=0.29$, with the strongest lines corresponding to H$\beta$, the [O III] doublet, and H$\alpha$. 

On the other hand, the redshift identification is more challenging for object C, as only one emission line is conspicuous. We consider the possibilities of this line being redshifted H$\alpha$, H$\beta$, or [O II]. We reject the first two propositions on the basis that it is highly improbable for only one of either H$\alpha$ and H$\beta$ to be strong, and that the remaining Balmer line should lie within the wavelength coverage of the spectrum in both cases. Thus, we conclude that this line is [O II], and that the object is an emission-line galaxy at $z=1.11$. In addition, a close examination of the spectrum reveals possible Ca H+K absorption lines, although they are weak. 

To sum up, objects A and C have dissimilar spectra with emission lines at different wavelengths, demonstrating that the two objects do not originate from a single source. Therefore, we conclude that we have discovered two new galaxies and, in turn, HULQ J0002+0239 is not a QSO lens, in contrast to our expectations. 

Both sources have magnitudes similar to the magnitude limit of Gemini Observations of Galaxies in Rich Early Environments (GOGREEN; \citealt{Balogh+17}), which is one of the deepest spectroscopic surveys being conducted with the Gemini Telescopes ($z < 24.25$ mag). The contrasting factor is that this study focuses on objects with emission lines, whereas GOGREEN aims to measure redshifts for galaxies. Our success in the redshift identification of these sources reiterates that deep GMOS spectroscopy of $\sim$5 hours is capable of measuring the redshifts of sources with emission lines as faint as $z \sim 24.5$ mag. 

\subsection{Nature of object A}
\begin{figure}
\begin{minipage}[c]{0.75\textwidth}
\includegraphics[width=\textwidth]{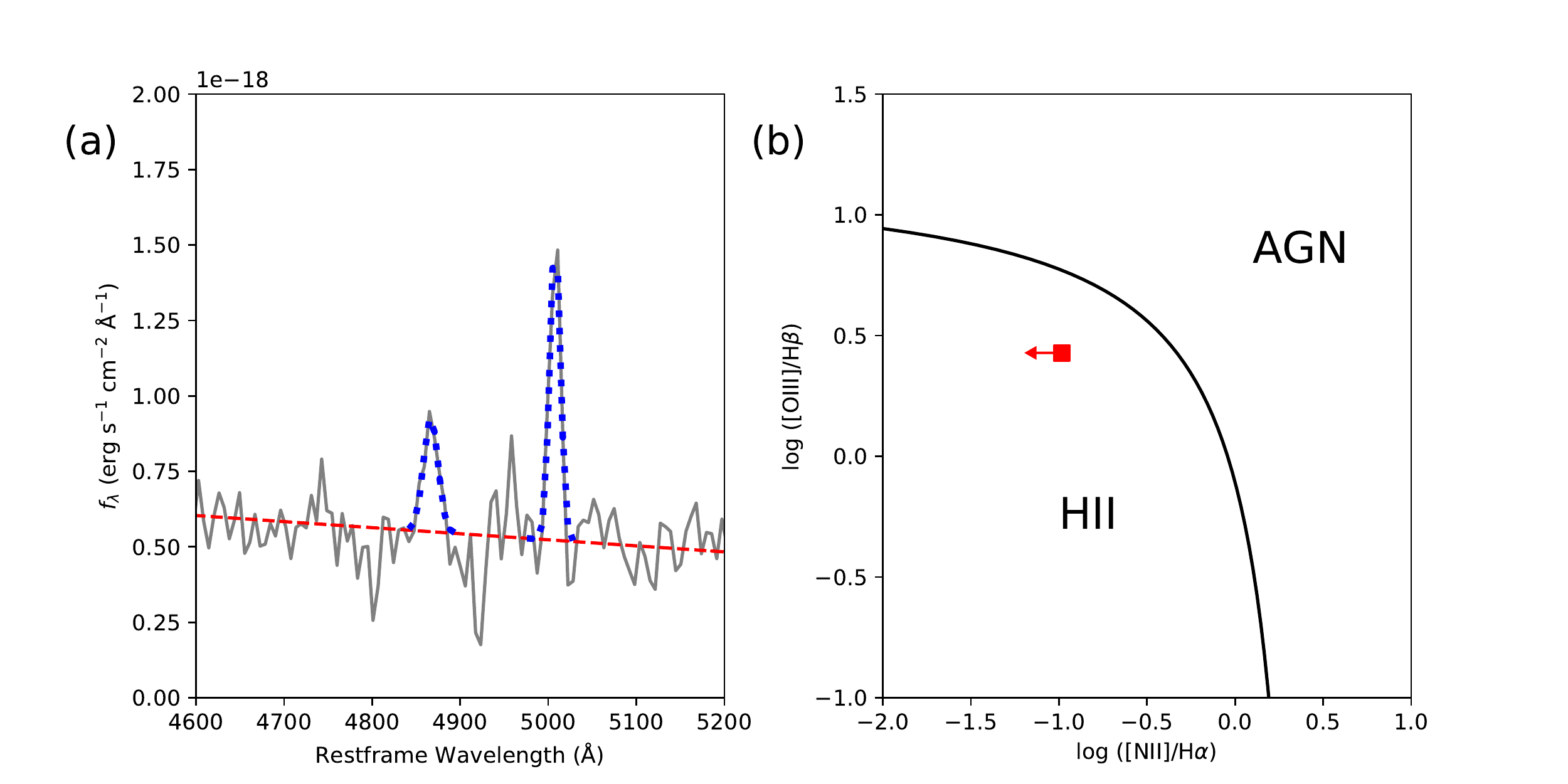}
\end{minipage}%
\begin{minipage}[c]{0.25\textwidth}
\caption{Spectral analysis of object A.
(a) Fitting results for the H$\beta$-region spectrum of object A. The spectrum is shown in gray, and the continuum and single-Gaussian emission line fits are shown with red dashed and blue dotted lines, respectively. (b) BPT diagram, with the line separating star-forming regions and AGNs from \citet{Kewley+06} shown in black. The red square represents object A, and the red arrow indicates the limit of the ratio. }\label{fig:bpt}
\end{minipage}
\end{figure}

The Baldwin-Philips-Terlevich (BPT) diagram \citep{Baldwin+81} is commonly used to distinguish AGNs from star-forming regions based on their emission line flux ratios. To measure the line fluxes, we used a power-law function to fit the continuum, and single-Gaussian functions were used to fit each of the H$\beta$, [O III]$\lambda$5007, and H$\alpha$ emission lines. Figure \ref{fig:bpt}(a) shows the fitting results for the H$\beta$ region of object A. 
The full widths at half maximum (FWHMs) of H$\beta$, [O III]$\lambda$5007, and H$\alpha$ are 1200 $\pm$ 200 km s$^{-1}$, 840 $\pm$ 100 km s$^{-1}$, and 730 $\pm$ 80 km s$^{-1}$, respectively. The similarity of the FWHMs of [O III]$\lambda$5007 and H$\alpha$ implies that a strong broad line component is absent, and the different FWHMs for H$\beta$ are mostly due to errors of the spectrum. Also, since the lines close to H$\alpha$ (e.g., [N II]$\lambda$6584, [S II]$\lambda$6717,6731, and [O I]$\lambda$6300) are very weak if at all existent, we were unable to measure these ratios accurately, and only placed upper limits on these ratios. Using these ratios and limits, object A is plotted on the BPT diagram in Figure \ref{fig:bpt}(b). Its position indicates that object A is a star-forming galaxy. This is also in agreement with the blue SED of object A; the restframe $g-r$ color obtained from the spectrum is $-0.05$, which places it far from the red sequence and reinforces the hypothesis of it being a star-forming galaxy. 

\subsection{Spectra from other objects}
\begin{figure}
\makebox[\textwidth][c]{\includegraphics[width=1.2\textwidth]{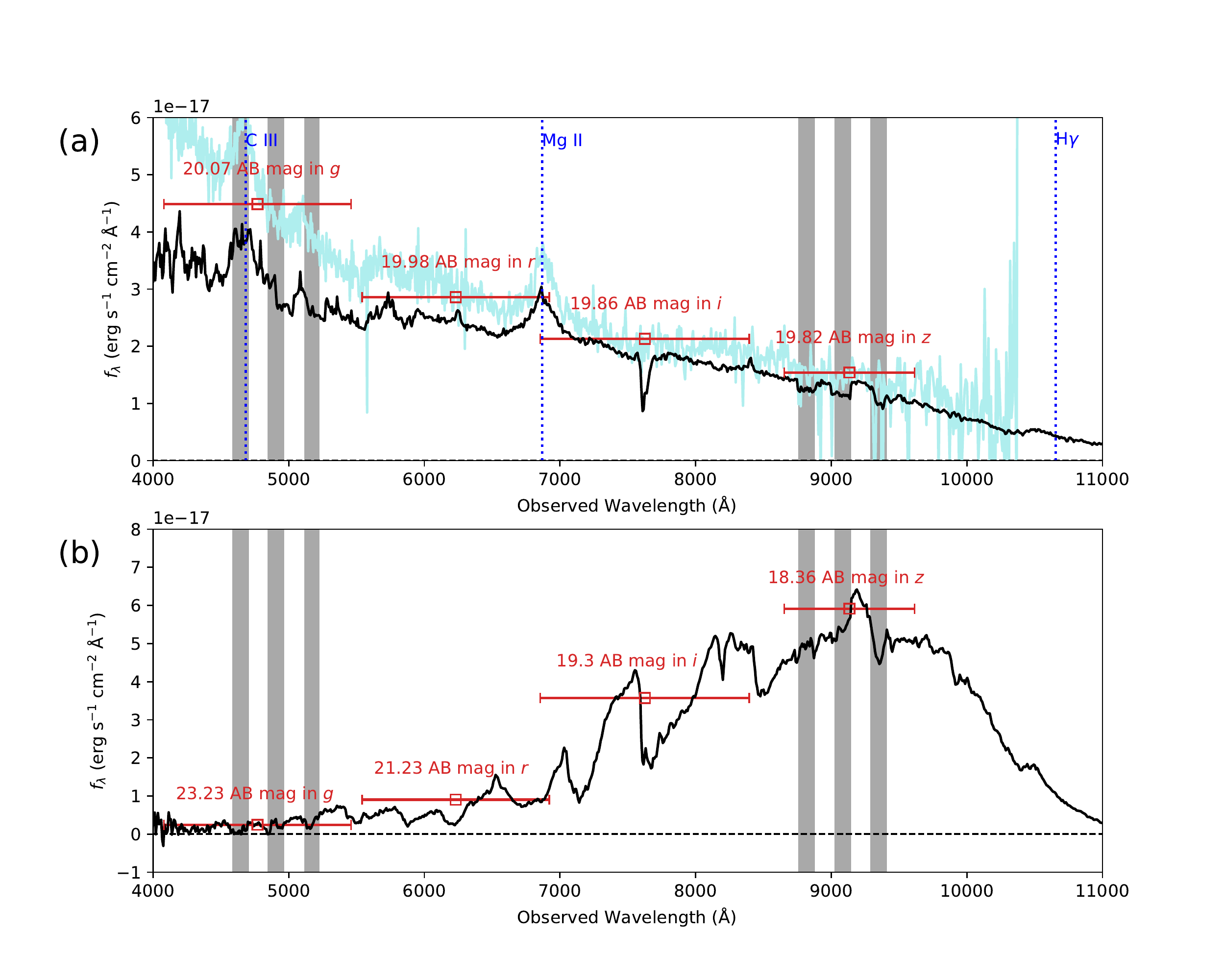}}
\caption{Gemini-N/GMOS spectra of SDSS J0002+0239 and SDSS J000216.35+023840.9.
(a) Spectrum of SDSS J0002+0239. The GMOS spectrum is shown in black, and the SDSS spectrum is shown in turquoise. The vertical blue dashed lines indicate the wavelengths of the emission lines for $z$=1.455, and the dark gray regions indicate the CCD gaps. Its magnitudes from SDSS are shown as red squares, with the error bars indicating filter widths. (b) Spectrum of SDSS J000216.35+023840.9, a star located $\sim$60$\arcsec$ southwest from the QSO. The horizontal black dashed line indicates zero flux. The dark gray regions, red squares, and the red error bars are identical to (a). 
}
\label{fig:spec2}
\end{figure}

We have obtained the spectrum of SDSS J0002+0239, which is shown in Figure \ref{fig:spec2}(a). It agrees well with the SDSS spectrum, and the slight flux decrease over all wavelengths can be attributed to the QSO being located slightly off-center within the slit. 

Besides the three objects that were intended to be within the slit, there are additional objects whose spectra were taken unintentionally at various positions along the slit. One notable object is SDSS J000216.35+023840.9, which is an object classified as ``STAR'' in the SDSS archives, and located $\sim$60$\arcsec$ southwest of SDSS J0002+0239. It has not been observed spectroscopically before, and its first spectrum is shown in Figure \ref{fig:spec2}(b). The photometric data from SDSS agree well with the spectrum, and it resembles that of a late-type star, most likely an M-type star.

\section{Implications for the QSO lens number density in HSC/Wide}
We discuss the statistical implications of HULQ J0002+0239 being a nonlens. To determine whether this is in line with expectations from Paper I, we estimate the number of large-$\theta_{\rm Ein}$ QSO lenses that should lie within the HSC/Wide PDR2 coverage, and compare this with our results. 

For this argument the areal coverage of HSC/Wide PDR2 is required. This is complicated to measure, but we attempted to estimate this using the number of SDSS QSOs observed in the four broadband filters, $g$, $r$, $i$, and $z$. The SDSS DR14 QSO catalog \citep{Paris+18} contains $\sim$530,000 QSOs over an area of $\sim$9,400 deg$^2$, resulting in an average surface density of 56 QSOs per square degree. The number of QSOs observed by all four filters in HSC/Wide PDR2 is $\sim$26,000, so we estimate the area observed by all four filters to be $\sim$470 deg$^2$. Because the DR14 QSO catalog was an intermediate release of an ongoing survey, the QSOs are not uniformly spread out over the survey area, but this factor should be negligible for the purpose of this discussion. 

We attempted another approach to measure the areal coverage. We created a two-dimensional surface that is representative of the celestial sphere, and divided it into identical bins in both RA and Dec directions. If a bin has at least one SDSS QSO that is observed in all four filters within it, it is regarded as occupied, and the area of all occupied bins summed up should be equivalent to the HSC/Wide PDR2 areal coverage. The size of the bins is critical in determining the areal coverage, and we selected 0.5 degrees in both directions to be the most appropriate. The areal coverage of HSC/Wide PDR2 based on this estimate is $\sim$680 deg$^2$. 

The two estimates are somewhat in conformity with each other, especially for our purpose of roughly estimating the number of QSO lenses. Therefore, we consider the median of these two values of 570 deg$^2$ to be the areal coverage of HSC/Wide PDR2. 

\begin{figure}
\includegraphics[width=0.5\textwidth]{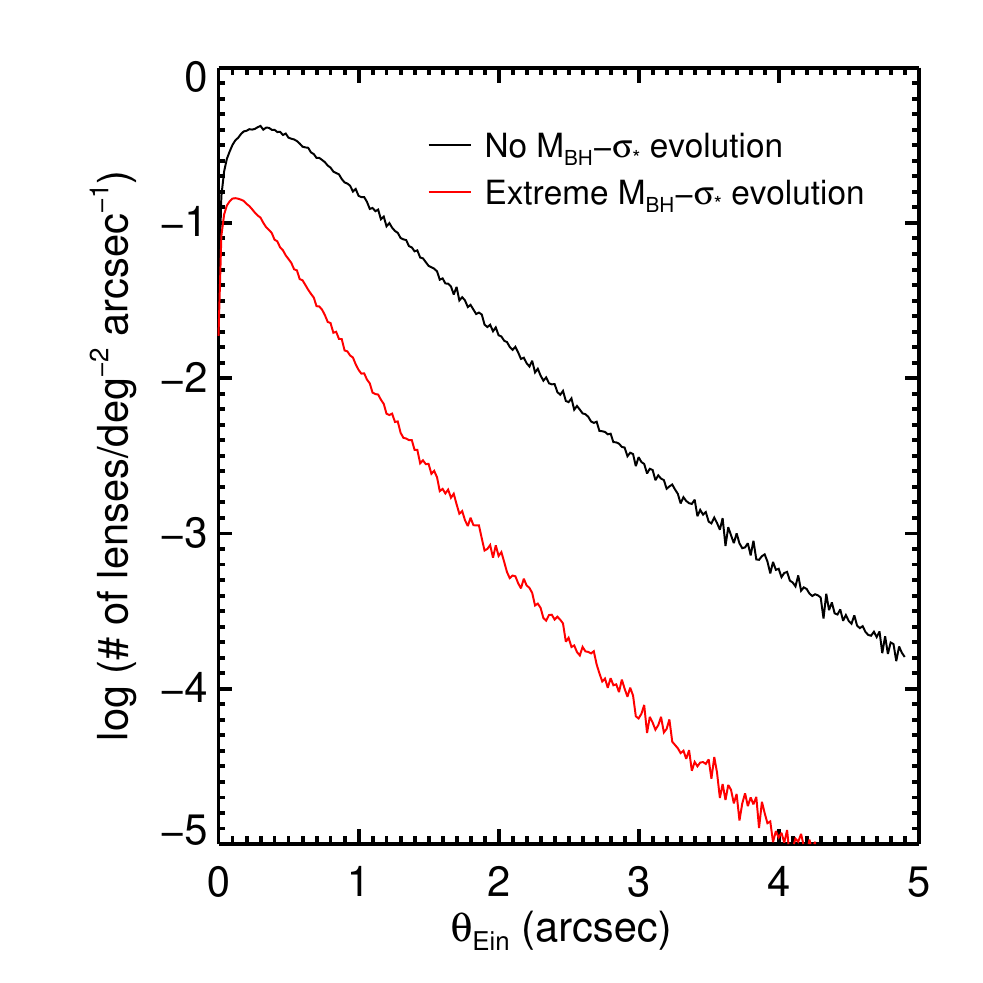}
\caption{$\theta_{\rm Ein}$ distribution of QSO lenses for the limiting magnitude of HSC/Wide. Black and red lines denote the distributions for the no-evolution scenario and when there is an extreme evolution of the $M_{\rm BH} - \sigma_*$ relation, respectively. This figure is a rescaled version of Figure 13 from \citet{TaakY+20}. 
}
\label{fig:rein}
\end{figure}

There are two scenarios regarding the evolution of the $M_{\rm BH} - \sigma_*$ relation: one where there is no evolution of the local relation \citep{Kormendy+13}, and the other for an extreme-evolution case \citep{WooJ+08}. Figure \ref{fig:rein} shows the $\theta_{\rm Ein}$ distribution of QSO lenses for the HSC/Wide depth for the two scenarios. Assuming that the shape of this distribution does not change for the effective depth of $i=25$ mag that was used for the QSO lens selection, we can calculate the expected number of QSO lenses with $\theta_{\rm Ein} > 3\farcs5$ to be 7.2$\times$10$^{-4}$ deg$^{-2}$ and 1.5$\times$10$^{-5}$ deg$^{-2}$ for the no-evolution and the extreme-evolution scenarios, respectively. Multiplying these by the areal coverage of HSC/Wide PDR2 gives 0.41 and 0.0089 QSO lenses, which implies that the existence of a QSO lens with $\theta_{\rm Ein} > 3\farcs5$ in HSC/Wide PDR2 is plausible for the no-evolution scenario, but unlikely for the extreme-evolution one. Thus, the rejection of HULQ J0002+0239 as a QSO lens is consistent with the expectations. 

HULQ J0002+0239 was the most promising large-separation QSO lens candidate in this study, and was one of two Grade A candidates with $\theta_{\rm Ein} > 3\farcs5$. Such systems require large $\theta_{\rm Ein}$, which imply group-scale deflectors and are intrinsically less common than galaxy-scale deflectors. This suggests that lens systems with $\theta_{\rm Ein} > 3\farcs5$ are likely to be spurious, and more focus should be given to small-$\theta_{\rm Ein}$ QSO lenses. 

We also calculated the expected number density of QSO lenses with smaller $\theta_{\rm Ein}$ in the range of $1\farcs5 < \theta_{\rm Ein} < 3\farcs5$, which is 2.6$\times$10$^{-2}$ deg$^{-2}$ and 9.9$\times$10$^{-4}$ deg$^{-2}$ for the two evolution scenarios. This results in 0.57--15 QSO lenses for the HSC/Wide PDR2 coverage, which is consistent with the 4+8 Grade A and B candidates from this study. We are currently in the process of improving the search mechanism to find these candidates, which will allow us to place meaningful constraints on the evolution of scaling relations. Furthermore, we plan to extend the search to future survey data, such as more recent HSC/Wide releases and imminent surveys conducted with newly built telescopes such as the Vera C. Rubin Observatory. This would increase the number of QSO lenses by a factor of several, and this will provide tighter constraints on the scaling relation evolution with increased number statistics.

\section{Conclusion}
We present a QSO lens search, using the HSC/Wide PDR2 imaging data, and the spectroscopic observation of HULQ J0002+0239, which is a promising QSO lens candidate resembling an Einstein cross. 
We obtained spectra of two lensed-object candidates, objects A and C, and measured their redshifts using prominent emission lines. However, their different redshifts at $z=0.29$ and $1.11$ imply that A and C are two newly discovered galaxies, and that HULQ J0002+0239 is not a QSO lens. In this process, we demonstrated that determining the redshifts of $z \sim 24.5$ mag sources with emission lines is achievable with exposure times of $\sim$5 hours. 

The number of QSO lenses with large $\theta_{\rm Ein}$ ($ > 3\farcs5$) is expected to be less than 1 within HSC/Wide PDR2, meaning that the null discovery of such large-$\theta_{\rm Ein}$ QSO lenses is in agreement with the predictions from Paper I. The number of QSO lenses with smaller $\theta_{\rm Ein}$ ($1\farcs5 < \theta_{\rm Ein} < 3\farcs5$) is anticipated to be 0.57--15 for HSC/Wide PDR2, depending on the evolution of the $M_{\rm BH} - \sigma_*$ relation, and this is consistent with the number of candidates from these data (12). Future confirmation of these candidates will help us discern the more plausible evolution scenario. The investigation of ongoing and upcoming large-area surveys with improved depth, areal coverage, and seeing, as well as improvements in the lens search methodology, should reveal more QSO lenses, and help us constrain the coevolution of SMBHs and their host galaxies across cosmic time.

\begin{acknowledgements}

This research is supported by the National Research Foundation of Korea (NRF) grants (No. 2020R1A2C3011091 and No. 2021M3F7A1084525) funded by the Korean government (MSIT). Y. C. T. is supported by the Basic Science Research Program through the National Research Foundation of Korea (NRF) funded by the Ministry of Education (grant No. 2021R1A6A3A14044070). Y. K. is supported by the National Research Foundation of Korea (NRF) grant funded by the Korean government (MSIT) (No. 2021R1C1C2091550) and acknowledges the support from the China Postdoc Science General (2020M670022) and Special (2020T130018) Grants funded by the China Postdoctoral Science Foundation. M. H. acknowledges the support from the Korea Astronomy and Space Science Institute grant funded by the Korean government (MSIT) (No. 2022183005).

Funding for the Sloan Digital Sky Survey IV has been provided by the Alfred P. Sloan Foundation, the U.S. Department of Energy Office of Science, and the Participating Institutions. SDSS-IV acknowledges support and resources from the Center for High-Performance Computing at the University of Utah. The SDSS web site is www.sdss.org. SDSS-IV is managed by the Astrophysical Research Consortium for the Participating Institutions of the SDSS Collaboration including the Brazilian Participation Group, the Carnegie Institution for Science, Carnegie Mellon University, the Chilean Participation Group, the French Participation Group, Harvard-Smithsonian Center for Astrophysics, Instituto de Astrof\'isica de Canarias, The Johns Hopkins University, Kavli Institute for the Physics and Mathematics of the Universe (IPMU) / University of Tokyo, the Korean Participation Group, Lawrence Berkeley National Laboratory, Leibniz Institut f\"ur Astrophysik Potsdam (AIP),  Max-Planck-Institut f\"ur Astronomie (MPIA Heidelberg), Max-Planck-Institut f\"ur Astrophysik (MPA Garching), Max-Planck-Institut f\"ur Extraterrestrische Physik (MPE), National Astronomical Observatories of China, New Mexico State University, New York University, University of Notre Dame, Observat\'ario Nacional / MCTI, The Ohio State University, Pennsylvania State University, Shanghai Astronomical Observatory, United Kingdom Participation Group,Universidad Nacional Aut\'onoma de M\'exico, University of Arizona, University of Colorado Boulder, University of Oxford, University of Portsmouth, University of Utah, University of Virginia, University of Washington, University of Wisconsin, Vanderbilt University, and Yale University. 

The Hyper Suprime-Cam (HSC) collaboration includes the astronomical communities of Japan and Taiwan, and Princeton University. The HSC instrumentation and software were developed by the National Astronomical Observatory of Japan (NAOJ), the Kavli Institute for the Physics and Mathematics of the Universe (Kavli IPMU), the University of Tokyo, the High Energy Accelerator Research Organization (KEK), the Academia Sinica Institute for Astronomy and Astrophysics in Taiwan (ASIAA), and Princeton University. Funding was contributed by the FIRST program from Japanese Cabinet Office, the Ministry of Education, Culture, Sports, Science and Technology (MEXT), the Japan Society for the Promotion of Science (JSPS), Japan Science and Technology Agency (JST), the Toray Science Foundation, NAOJ, Kavli IPMU, KEK, ASIAA, and Princeton University. This paper makes use of software developed for the Large Synoptic Survey Telescope. We thank the LSST Project for making their code available as free software at  http://dm.lsst.org. The Pan-STARRS1 Surveys (PS1) have been made possible through contributions of the Institute for Astronomy, the University of Hawaii, the Pan-STARRS Project Office, the Max-Planck Society and its participating institutes, the Max Planck Institute for Astronomy, Heidelberg and the Max Planck Institute for Extraterrestrial Physics, Garching, The Johns Hopkins University, Durham University, the University of Edinburgh, Queen's University Belfast, the Harvard-Smithsonian Center for Astrophysics, the Las Cumbres Observatory Global Telescope Network Incorporated, the National Central University of Taiwan, the Space Telescope Science Institute, the National Aeronautics and Space Administration under Grant No. NNX08AR22G issued through the Planetary Science Division of the NASA Science Mission Directorate, the National Science Foundation under Grant No. AST-1238877, the University of Maryland, and Eotvos Lorand University (ELTE) and the Los Alamos National Laboratory. Based [in part] on data collected at the Subaru Telescope and retrieved from the HSC data archive system, which is operated by Subaru Telescope and Astronomy Data Center at National Astronomical Observatory of Japan.

This paper includes data gathered with the 6.5 meter Magellan Telescopes located at Las Campanas Observatory, Chile. 

This work was supported by the K-GMT Science Program (PID: GN-2020B-Q-213) of Korea Astronomy and Space Science Institute (KASI). 

Based on observations obtained at the international Gemini Observatory, a program of NSF's NOIRLab, which is managed by the Association of Universities for Research in Astronomy (AURA) under a cooperative agreement with the National Science Foundation. on behalf of the Gemini Observatory partnership: the National Science Foundation (United States), National Research Council (Canada), Agencia Nacional de Investigaci\'{o}n y Desarrollo (Chile), Ministerio de Ciencia, Tecnolog\'{i}a e Innovaci\'{o}n (Argentina), Minist\'{e}rio da Ci\^{e}ncia, Tecnologia, Inova\c{c}\~{o}es e Comunica\c{c}\~{o}es (Brazil), and Korea Astronomy and Space Science Institute (Republic of Korea).

This work was enabled by observations made from the Gemini North telescope, located within the Maunakea Science Reserve and adjacent to the summit of Maunakea. We are grateful for the privilege of observing the Universe from a place that is unique in both its astronomical quality and its cultural significance. 

IRAF is distributed by the National Optical Astronomy Observatory, which is operated by the Association of Universities for Research in Astronomy (AURA) under a cooperative agreement with the National Science Foundation.

PyRAF is a product of the Space Telescope Science Institute, which is operated by AURA for NASA. 

Data reduction for the GMOS data were conducted following the GMOS Data Reduction Cookbook (Shaw, R. A. 2016, GMOS Data Reduction Cookbook (Version 1.1; Tucson, AZ: National Optical Astronomy Observatory)).

\end{acknowledgements}

\bibliographystyle{aa}
\bibliography{main.bbl}

\clearpage

\clearpage

\clearpage

\clearpage

\clearpage

\clearpage

\clearpage

\clearpage

\clearpage

\begin{table}
\caption{Photometric information for the objects}
\label{tbl:phot}
\centering
\begin{tabular}{c c c c c c}
\hline\hline
Object  & IMACS $u'$& HSC $g$   & HSC $r$   & HSC $i$   & HSC $z$ \\
        & (mag)     & (mag)     & (mag)     & (mag)     & (mag) \\
\hline
A  & $>$24.9\tablefootmark{a} & 25.54 $\pm$ 0.18        & 25.00 $\pm$ 0.11  & 24.91 $\pm$ 0.15 & 25.18 $\pm$ 0.32 \\
B  & $>$24.9\tablefootmark{a} & $>$26.6\tablefootmark{a}        & $>$26.2\tablefootmark{a}      & 25.33 $\pm$ 0.22 & $>$25.3\tablefootmark{a}\\
C  & $>$24.9\tablefootmark{a} & 25.57 $\pm$ 0.19        & 25.24 $\pm$ 0.13  & 24.71 $\pm$ 0.13 & 24.68 $\pm$ 0.21 \\
D  & $>$24.9\tablefootmark{a} & $>$26.6\tablefootmark{a}& 25.72 $\pm$ 0.22  & 25.54 $\pm$ 0.26 & $>$25.3\tablefootmark{a}\\
\hline
\end{tabular}
\tablefoot{\tablefoottext{a}{In the case of non-detections, the 5$\sigma$ limiting magnitude for point sources is shown instead.}}
\end{table}

\begin{table}
\caption{Model parameters used for the \texttt{lensmodel} simulation}
\label{tbl:lensmodel}
\centering
\begin{tabular}{c c}
\hline\hline
Parameter & Value \\
\hline
\multicolumn{2}{c}{\textit{Singular Isothermal Ellipsoid}}\\
Einstein radius ($\theta_{\rm Ein}$)                    & 3$\farcs$56 \\
Ellipticity                                             & 0.157\\
Position angle (east of north)  & -35.0$^{\circ}$ \\ 
\hline
\multicolumn{2}{c}{\textit{Source}}\\
Position (relative to QSO)                      & (0$\farcs$0031, -0$\farcs$0036)\\
\hline
\end{tabular}
\end{table}

\begin{table}
\caption{$i$-band observed and simulated flux ratios of the objects}
\label{tbl:lensmodel2}
\centering
\begin{tabular}{c c c}
\hline\hline
Object  & $F_i/F_A$ (observed)  & $F_i/F_A$ (simulated)\\
\hline
A & 1.000 $\pm$ 0.137 & 1.016\\
B & 0.678 $\pm$ 0.139 & 0.760\\
C & 1.194 $\pm$ 0.139 & 0.816\\
D & 0.559 $\pm$ 0.135 & 0.822\\
\hline
\end{tabular}
\end{table}

\end{document}